\documentclass[aps,pra,twocolumn,groupedaddress,showpacs]{revtex4}
\usepackage{times}
\usepackage{graphicx}
\usepackage{amsmath}

\newcommand{\ket}[1]{|#1\rangle}

\newcommand{\adj}[1]{{#1}^\dag}
\newcommand{\deldelsq}[2]{\frac{\partial^2#1}{\partial{#2}^2}}
\newcommand{\pd}[2]{\frac{\partial#1}{\partial#2}}
\newcommand{\conj}[1]{{#1}^\ast}
\newcommand{\rme}{{\mathrm e}}
\newcommand{\abs}[1]{\modulus{#1}}
\newcommand{\modulus}[1]{\left|#1\right|}

\newcommand{\eplus}[0]{e_+(\xi,\tau)}
\newcommand{\conjeplus}[0]{\conj{e}_+(\xi,\tau)}
\newcommand{\eminus}[0]{e_-(\xi,\tau)}
\newcommand{\conjeminus}[0]{\conj{e}_-(\xi,\tau)}
\newcommand{\wfm}[0]{\psi_m(\xi,\tau)}
\newcommand{\wfmminus}[0]{\psi_{m-2}(\xi,\tau)}
\newcommand{\wfmplus}[0]{\psi_{m+2}(\xi,\tau)}
\newcommand{\conjwfmminus}[0]{\conj{\psi}_{m-2}(\xi,\tau)}
\newcommand{\conjwfmplus}[0]{\conj{\psi}_{m+2}(\xi,\tau)}

\newcommand{\pdxi}[1]{\pd{#1}{\xi}}
\begin{document}
%\preprint{}

\title{Rayleigh superradiance and dynamic Bragg gratings in an end-pumped Bose-Einstein condensate}
\author{A. Hilliard}\email[]{hilliard@nbi.dk}
\author{F. Kaminski}
\author{R. le Targat}
\author{C. Olausson}
\author{E.S. Polzik}
\author{J. H. M\"uller}
\affiliation{Niels Bohr Institute, Blegdamsvej 17, 2100 Copenhagen {\O}, Denmark}

\date{\today}

\begin{abstract}
We study experimentally superradiant Rayleigh scattering from a Bose-Einstein condensate (BEC) in a new parameter regime  where pump depletion and the exchange of photons between the endfire modes are important.
Through experiments and simulations we show that
collective atom light coupling leads to the self-organized
formation of dynamic Bragg gratings within the sample. These gratings
lead to an efficient back-scattering of pump photons and optical
resonator structures within the BEC.
\end{abstract}
% insert suggested PACES numbers in braces on next line
%\keywords{}
\maketitle
With its extremely high optical depth and unique coherence
properties, a Bose Einstein condensate provides an ideal object
on which to study collective light scattering with the goal of
generating and probing light-atom correlations.
Superradiant light scattering (SLS) from ultra-cold atomic
ensembles has been recognized as a method to generate
entangled atoms and photons due to the fact that the
interaction Hamiltonian has the generic form of a parametric
amplifier $H\propto \adj{\hat{a}} \adj{\hat{b}}$, where
correlations based on momentum conservation arise in the case
of Rayleigh scattering \cite{PhysRevA.70.043809}, and angular
momentum in the case of Raman processes
\cite{PhysRevLett.85.5026}. Entanglement in the ultra-low gain regime of such an interaction, when far less than one atom-photon excitation pair is generated on average, forms the basis for a quantum repeater \cite{DLCZ_Nature} and has been  studied extensively \cite{Kimble_Qinternet}. Here, we are interested in the high gain - superradiant - regime where the detection of entanglement
requires in general the measurement of both the phase and
amplitude of the light and matter waves at the sub-shot noise
level. While this is typically hampered by the lack of an
atomic analogue to homodyne detection of light, a recent
proposal claims entanglement may be detected  by just counting photons and atoms \cite{Muschik}.
Nonetheless, the full dynamics of coupled matter and light
waves where one does not place limitations on the depletion of
either the condensate or input light, and where propagation
effects are considered, remain important issues if such
correlations and entanglement are to be made useful resources.

To this end, we explore superradiant Rayleigh scattering from a
trapped, cigar shaped BEC as the pump detuning is varied while the single
particle scattering rate $R$ is kept constant. In this way, we
investigate the effect of the detuning of the pump beam in the
process, and move between the case where the pump beam remains
essentially undepleted by the scattering, to the situation
where superradiant scattering is `clamped' by a lack of photons
in the pump beam. Crucial to these dynamics is the structure
that builds up along the long axis of the condensate,
demonstrating characteristics from `Dicke' superradiance from
extended samples \cite{PhysRevA.14.1169,SR_review_GrossHaroche,Zheleznyakov_review}, a fact recognized
recently both experimentally
\cite{DominikSchneble04182003,sadler:110401} and theoretically
\cite{zobay:041604,PhysRevA.73.013620,uys:033805,Avetisyan_JETP_130_771_2006}. Contrary to most
earlier experimental work on the subject
\cite{S.Inouye07231999,DominikSchneble04182003,PhysRevA.69.041601,yoshikawa:041603},
we end-pump along the long axis with a beam mode-matched to the transverse cross-section of the BEC
in order to optimize the coupling between pump photons and
atoms.
As opposed to the study in \cite{Li_PhysLettA}, which also probed the BEC along its long axis, we use the back-scattered light to monitor the superradiant dynamics with very good time resolution.
In view of our ultimate goal to probe light-atom correlations, we have performed a careful calibration of the detection system. As a consequence, we find quantitative agreement with simulations of a 1D semi-classical model over a wide parameter range.
The quality of the agreement with experiment supports the use of the simulations to understand the spatial and temporal dynamics inside the sample.

\begin{figure}
 \includegraphics[width=8cm]{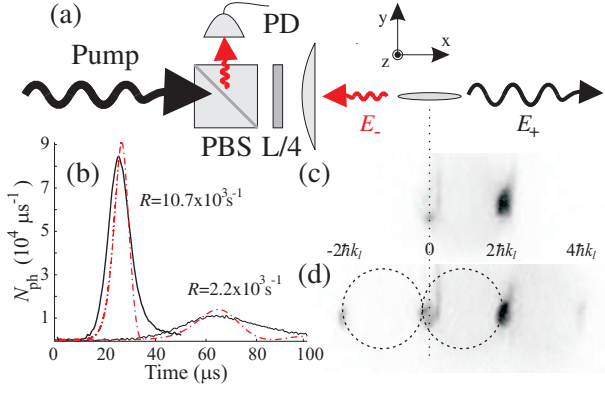}%
 \caption{(Color online) (a) %Schematic diagram of the experiment.
 The BEC was end-pumped by a focused, circularly polarized laser beam,
 and the back-scattered light was directed through a quarter wave plate (L/4)
 and polarizing beam-splitter (PBS) onto a sensitive
 photodetector (PD). (b) Traces from PD are shown for low and high pump
 powers
 %corresponding to single particle Rayleigh scattering rates
% of
% $R=2.2\times10^3$s$^{-1}$ and $15.9\times 10^3$s$^{-1}$ respectively
 at a detuning of
 \mbox{$\delta=-2\pi\times2.6$~GHz}. Simulations for the same parameters are presented as
 red dash-dot lines. Absorption images after 45~ms time of flight (TOF) of the corresponding
 atomic distributions are shown in (c) (low power, pump pulse duration 200~$\mu$s) and (d) (high power,  pump pulse duration 50~$\mu$s). Circles indicating the separation of adjacent momentum orders after 45~ms TOF are shown in (d); note that they originate from the input light end ($x=0$) of the BEC.
 \label{fig:figure_1}}
 \vspace{-0.5cm}
 \end{figure}

Like superradiance (SR) in electronically
inverted samples \cite{PhysRev.93.99}, superradiant light scattering from an ultra-cold atomic sample is a process whereby an initially
unoccupied electromagnetic field mode becomes weakly populated
through spontaneous emission; these modes are then amplified with
the highest gain along the direction of greatest optical depth.
In the end-pumped geometry (see Fig.
\ref{fig:figure_1}), photons are absorbed from the forward
travelling beam $E_+$ and scattered into the backward
travelling beam $E_-$, leading to a concomitant change in the
scattered atom's momentum of $2\hbar k_l$ and a recoil shift in
the back-scattered light of four times the recoil frequency
$\omega_r$. The forward and backward travelling light waves
interfere to give a moving intensity grating with a spatial
period of $\lambda/2$ that varies in visibility and phase over
the length of the sample. In the case of Rayleigh scattering,
the internal ground state of the atom does not change in the
process, leading to a density modulation due to interference
between the different momentum orders that has the same period
as the light intensity modulation. Thus, the physical picture
is one of four wave-mixing.

In our experiments, superradiant Rayleigh scattering was induced in a trapped BEC by
illuminating it with a pulse of off-resonant light along the long
axis of the condensate, as shown in Fig.~\ref{fig:figure_1}. The BEC was
generated by evaporatively cooling a cloud of $^{\textrm{\small{87}}}$Rb atoms in the
$\ket{F=1,m_F=-1}$ hyperfine state in a Ioffe-Pritchard magnetic trap.
We obtain cigar shaped condensates containing $1.35\times10^6$
atoms, with in-trap Thomas Fermi radii of \mbox{$r_\bot=6.4$} and \mbox{$r_\|=65$ $\mu$m} in
the radial and axial directions and with no discernible thermal
fraction. The pump light was detuned by a variable amount from
the $\ket{F=1,m_F=-1}\rightarrow \ket{F=2,m_F=-2}$ transition
on the D1 line of $^{\textrm{\small{87}}}$Rb at 795 nm, and
circularly polarized with respect to the long axis of the trap.
All data presented is for red detunings
($\delta=\omega_l-\omega_0 < 0$); rectangular pump pulse
envelopes; and where the atoms were interrogated in-trap, with
the trapping potential extinguished immediately after the end
of the pump pulse. The beam was focused to a waist of
%13.2$\mu$m
13~$\mu$m at the center of the condensate with negligible
change of beam size over the length of the BEC. Light was back-scattered  by the sample in the same polarization as the input
beam, and thus the backward travelling light was reflected by
the polarizing beamsplitter, then directed onto a sensitive PIN diode
photodetector. The detector, with a bandwidth of 400~kHz, is
shot-noise limited for photon fluxes greater than $10^5~\mu$s$^{-1}$. To avoid back reflections from optics and cell
windows that seed the process, the pump beam was inserted at a
slight angle (less than $2^\circ$). Pictures of the atoms were
obtained after 45~ms time of flight by resonant absorption
imaging.
Figure \ref{fig:figure_1}(b) shows experimental and simulated time traces for high and low pump powers, with
the corresponding atomic distributions shown in Fig.
\ref{fig:figure_1}(c) and (d). For single particle scattering
rates  much smaller than the recoil frequency
$\omega_r=2\pi\times3.6$~kHz, scattering to higher atomic
momentum orders occurs sequentially on a time scale \mbox{$\sim \tau_r=2\pi/\omega_r$.}
%Footnote: There is not necessarily a one-to-one correspondence between
%multiple light pulse emission and cascading transfer to higher momentum states
%\cite{PhysRevA.14.1169}.
% although it should be
%noted that subsequent pulses may be the result of `ringing'
%behaviour rather than a populating of a subsequent order.
%`Ringing' is a feature of superradiance in extended samples
%where several superradiant pulses may be emitted due to local
%growth in the coherences (the grating) due to excitation by
%other parts of the sample \cite{PhysRevA.14.1169}.
Figure~\ref{fig:figure_1}(c) shows this case, where the
transfer is limited to the first order. When $R\sim\omega_r$,
there is sufficient gain  for atoms to be back-scattered  into
negative momentum orders, i.e., the Kaptiza-Dirac regime where
atoms absorb back-scattered light and re-emit into the forward
direction, as is evident in Fig.~\ref{fig:figure_1}(d). An
 asymmetry in distance between forward and backward
scattered atoms and the center of the original condensate after time of
flight, observed previously in
\cite{DominikSchneble04182003,zobay:041604}, is visible in
Fig.~\ref{fig:figure_1}(d). This asymmetry can be traced
back to the spatial inhomogeneity of superradiant scattering
favoring the input end of the condensate, where the amplitude
of the reflected light $E_-$ is highest. The spatial dimensions
of the condensate and the slow expansion upon release from the
trap along the long axis are such that this spatial feature of
the scattering is evident after 45~ms time of flight. The momentum distributions are somewhat distorted due to the input angle of the beam: when the beam is aligned parallel to the long axis of the BEC, the Bragg condition is satisfied for the atoms centered around zero transverse momentum, but at the slight incident angle used, the patterns become more complicated. Nevertheless, the arrival times and amplitudes of the first superradiant light pulses are  very well described by the simulations.

The dynamics are simulated by numerically solving 1D
Maxwell-Schr\"odinger equations for the evolution of the atomic
wavefunctions and the incident and generated light fields
\cite{PhysRevA.14.1169,PhysRevA.20.2047}. The situation of SLS from a BEC was derived in \cite{zobay:041604,PhysRevA.73.013620}, and we follow closely their notation. In the end-pumped geometry, the  Maxwell-Schr\"odinger equations read:
\begin{equation}\label{eqn:SR_wf}
\begin{split}
i\frac{\partial\wfm}{\partial\tau} = & -\frac{1}{2}\deldelsq{\psi_m(\xi,\tau)}{\xi}
-i m \pd{\wfm}{\xi}\\
 & +\Lambda \conjeminus\eplus\wfmminus\rme^{2i(m-1)\tau}\\
 & +\Lambda\conjeplus\eminus\wfmplus\rme^{-2i(m+1)\tau}\\
 & +\Lambda(\abs{\eplus}^2+\abs{\eminus}^2)\wfm ,\raisetag{12pt}
\end{split}
\end{equation}
\begin{equation}\label{eqn:eplus}
\begin{split}
\pdxi{\eplus}& =  -i\Lambda \sum_{m=2n} \eminus\rme^{-2i(m-1)\tau}\\
                & \times \wfm\conjwfmminus + \eplus\abs{\wfm}^2,\raisetag{38pt}
\end{split}
\end{equation}
\begin{equation}\label{eqn:eminus}
\begin{split}
\pdxi{\eminus}& =  +i\Lambda \sum_{m=2n} \eplus\rme^{2i(m+1)\tau}\\
                 & \times\wfm\conjwfmplus + \eminus\abs{\wfm}^2,\raisetag{38pt}
\end{split}
\end{equation}
such that $\psi(x,t)=\sum_{m}\psi_m(x,t)\rme^{-i(\omega_m t-m k_l x)}$ is the slowly varying 1D wavefunction of momentum order \mbox{$m=2n$} for integer $n$, with $\omega_m=m^2\omega_r$; and the electric field (suppressing polarization) is given by \mbox{$E = E_+(x,t)\rme^{-i(\omega_l t- k_l x)}
+E_-(x,t)\rme^{-i(\omega_l t+ k_l x)}+ h.c.$}; where  $E_\pm(x,t)$ are slowly varying envelopes. The light fields are scaled such that \mbox{$E_\pm=e_\pm \sqrt{\hbar\omega_r k_l/\varepsilon_0 A}$}; and the
scaled space and time variables are related to real space parameters by $\xi=k_l x$ and $\tau=2\omega_r t$.
The coupling constant describing the strength of the interaction is given by \mbox{$\Lambda=1/4\cdot
\Gamma/\delta\cdot\sigma_0/A$}, where %\mbox{$\Lambda=\frac{1}{4\pi}\frac{\Gamma}{\delta}\frac{\sigma_0}{A}$}
 $\Gamma$ denotes the linewidth of the optical transition,
$\sigma_0$ the resonant atomic
absorption cross section, and $A$ the cross section of the
interaction region. Retardation effects have been neglected in
Eqns.~(\ref{eqn:eplus}) and (\ref{eqn:eminus}) given the length
of the condensate \mbox{$2r_\|=130~\mu$m}, which allows us to discard a
time derivative term.
In this approximation, these equations can be considered as describing the self-consistent light field distribution in a continuous array of Bragg gratings formed by the density modulation of the matter wave.
In this 1D model, the endfire mode described by $e_{+}$ and the pump mode coincide.
A 1D treatment of the problem is supported by the fact that the
Fresnel number of the condensate is of order one, which
suggests that the emission of light is predominantly into a
single transverse mode \cite{PhysRevA.3.1735}.
The harmonic trapping potential and the
mean field interaction term describing collisions between atoms
have been omitted, because they have a limited effect on the focus of these experiments - the first superradiant pulse \footnote{While a BEC is not
a prerequisite for superradiant scattering, it allows us here to neglect dephasing of density modulations due to thermal atomic motion.}.

Equations~(\ref{eqn:SR_wf}), (\ref{eqn:eplus}) and
(\ref{eqn:eminus}) describe a Raman interaction where a ladder
of momentum states is coupled by two counter-propagating light
fields. It is terms three and four that dominate in Eqn.~(\ref{eqn:SR_wf}): these
describe the local coupling
to the nearest momentum states via exchange of photons between
$e_+$ and $e_-$.
%The first two terms in Eqn. (\ref{eqn:SR_wf}) describe
%the quantum diffusion of the wave function and the momentum
%displacement induced by recoil respectively. For our parameters
%and interaction times, these terms contribute very little to
%the dynamics.
%terms three and four describe the local coupling
%to the nearest momentum states via exchange of photons between
%$e_+$ and $e_-$.
The final terms in Eqn. (\ref{eqn:SR_wf})
account for phase rotation of the matter wave due to the light
shift. Equations~(\ref{eqn:eplus}) and (\ref{eqn:eminus}) show
terms equivalent to the coupling terms in Eqn.
(\ref{eqn:SR_wf}). Specifically, the growth of $e_-$ occurs
with the corresponding growth of recoiling atoms $\psi_{m+2}$
and the decrease of $e_+$ photons and $\psi_m$ atoms. The last
terms in Eqns. (\ref{eqn:eplus}) and (\ref{eqn:eminus})
describe the effect on the light of the slowly varying refractive index due
to the large scale atomic density distribution.

To compare experimental results with simulations, the equations
(\ref{eqn:SR_wf}), (\ref{eqn:eplus}) and (\ref{eqn:eminus})
were solved numerically for experimental parameters. The
initial wavefunction $\psi_0$ was taken to be a 1D Thomas-Fermi
profile normalized to the number of atoms in the trap $N_\text{at}$.
The boundary conditions for the light fields were typically
taken to be $e_+(0,\tau)=e_{i}$ and $e_-(k_lL,\tau)=0$. Note that
$\abs{e_-(0,\tau)}^2$ is proportional to the
back-scattered light intensity measured in experiment, %$\clubsuit$\textbf{part added}
and that the total photon flux is conserved: $\abs{e_+(0,\tau)}^2=\abs{e_-(0,\tau)}^2+\abs{e_+(k_l L,\tau)}^2$.
As Eqns. (\ref{eqn:SR_wf}), (\ref{eqn:eplus}) and
(\ref{eqn:eminus}) contain no explicit noise term to instigate
superradiant scattering, we seed the process by taking a
non-zero first order momentum component
$\psi_{2}=\psi_0/\sqrt{N_\text{at}}$, corresponding to a single
delocalized  atom  in the first
side-mode \cite{PhysRevA.73.013620}. The random nature of the
initiation of superradiant scattering may be modeled in the
present formalism by using random initial conditions but here
we focus on the large scale dynamics of the system rather than
its noise properties \cite{PhysRevA.20.2047}.
 \begin{figure}[ht!]
 \includegraphics[width=8cm]{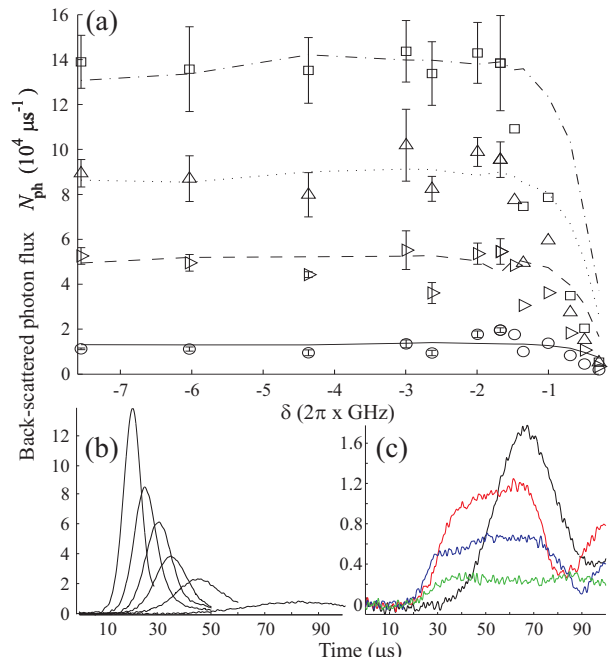}%
 \vspace{-0.4cm}
 \caption{(Color online) (a) Experimental (symbols) and simulated (lines) peak photon flux  of the first
          superradiant pulse as a function of detuning for
          $R=$2.2 (circles, solid line), 6.4 (right triangles, dashed line), 10.7
          (upward triangles, dotted line), 15.9 (squares, dash-dot line)$\times 10^3$~s$^{-1}$. Errorbar limits are the standard error of the mean of
          5 realizations, and for clarity are omitted in the low detuning portion of the graph.
          (b) Traces corresponding to the flat portion of Fig.~\ref{fig:2}(a).
          The detuning was $\delta=-2\pi\times2.6$~GHz, for  $R=2.2, 4.3, 6.4, 8.6, 10.7, 15.9\times10^{3}$~s$^{-1}$.
           For clarity, the traces are edited to
          show only the first pulse. (c) Traces corresponding to the
          decaying portion in Fig. \ref{fig:2}(a). For $R=2.2\times10^{3}$~s$^{-1}$ and
          detunings $\delta=-2\pi\times$ 2.0 (black), 1.0 (red), 0.7 (blue),
          0.5 (green)~GHz.
 \label{fig:2}}
 \vspace{-0.5cm}
 \end{figure}

Figure \ref{fig:2}(a) shows, for both experimental data (symbols) and simulations (lines), the peak values of the first superradiant pulse as the detuning of the pump beam was varied
while keeping the single particle scattering rate $R$ constant.
Results are shown for four values of $R$ in the range
$R\ll\omega_r$ to $R\sim\omega_r$. For a large portion of the
graph, the peak value is essentially independent of the
detuning, and there is excellent agreement between simulations and data. The input powers in the simulations have been scaled up by a common factor of 22\% \footnote{The main source of uncertainty in the coupling strength is the geometrical overlap between beam and BEC.}.
For lower detunings, the peak value of the emitted pulse falls
away, and the experimental data reaches our detection resolution for
$\delta\approx-2\pi\times 300$~MHz, while the simulations show the same qualitative behaviour.

To further study the two  regimes, Fig. \ref{fig:2}(b) shows experimental time traces
for several different values of $R$ at a high detuning, and
Fig. \ref{fig:2}(c) shows experimental traces for four low detunings while
$R$ is kept constant. Figure~\ref{fig:2}(b) confirms the
qualitative features evident in Fig.~\ref{fig:figure_1}(b): in the high detuning regime, the superradiant peaks arrive
earlier and are more sharply peaked the higher the pump power.
Simple models of superradiance predict a pulse height
proportional to $R$ and a pulse delay inversely proportional to
$R$ \cite{SR_review_GrossHaroche}, while we observe a faster increase of the peak power and a
slower decrease in pulse delay both experimentally and in the
simulations. Figure~\ref{fig:2}(c) shows how the superradiant
pulse intensity is clamped at low detunings.
The lower amplitudes can be traced
back to the lower incident photon fluxes and increasingly
important pump depletion in the low detuning case. The transition from
high detuning behaviour to the pump-depletion regime occurs in the experimental data
at points where the incident
photon flux is approximately 10 times the peak reflected photon
flux. Above this point, while the amplitude of the reflected
pulses drops, the observed peak reflectivity of the sample increases
sharply, up to values of $\sim30\%$. The simulations show the same qualitative behaviour in this regime, and we attribute the loss of quantitative agreement to the incoherent losses that are not captured by the model: i.e., emission into different modes. The spatial dependence of the light and matter waves plays a critical role here, and to gain more insight into the behavior of the system, we explore the light and matter
wave dynamics inside the sample through simulations.

 \begin{figure}
 \includegraphics[width=8cm]{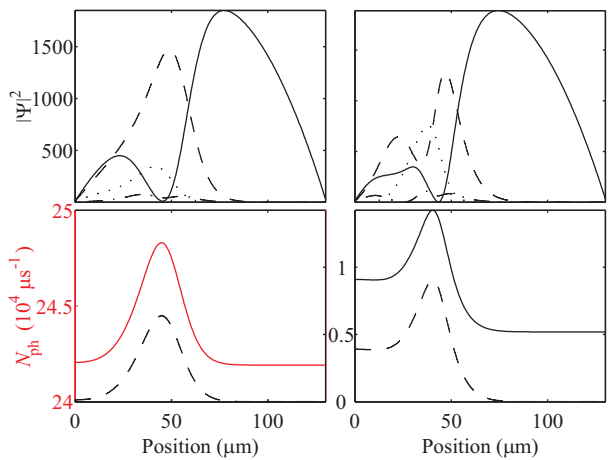}%
 %\vspace{-0.2cm}
 \caption{(Color online) Snapshots of system dynamics inside the sample for high detuning $\delta=-2\pi\times$ 2~GHz %$\clubsuit$\textbf{(Note:Changed from 2.6 GHz, Incorrectly labelled in original)}
          ((a) and (c)) and low detuning  $\delta=-2\pi\times$ 294 MHz
          ((b) and (d)) at the same rate  $R=2.2\times10^3$s$^{-1}$ after an interaction
          time $t=87~\mu$s.
          (a) and (b) show atomic densities in the momentum modes: $\abs{\psi_{-2}}^2$(dash-dot line), $\abs{\psi_0}^2$(solid line),
          $\abs{\psi_2}^2$(dashed line), $\abs{\psi_4}^2$(dotted line).
          (c) and (d) display light flux ($\propto \abs{e_{\pm}}^2$) within the sample
          for pump light $e_+$ (solid lines; in (c), red line and left axis)
          and reflected light $e_-$ (dashed lines; right scale in (c)). Note the shared scale of $e_-$ in (c) and both  $e_+$ and  $e_-$ in (d).
 \label{fig:3}}
 \vspace{-0.5cm}
 \end{figure}
Figure \ref{fig:3} shows the results of simulations for the lowest pump power case for
high (left side) and low (right side) pump detuning, after
87~$\mu$s of interaction time with the pump beam. Along the long axis of the BEC, atomic distributions are shown in the upper row ((a) and (b)), and the scaled intensities  $\propto\abs{e_+}^2$ and $\abs{e_-}^2$ in the lower row ((c) and (d)). The general dynamics for low input power, high detuning are as follows: The back-scattered light intensity in the sample builds up at the  input end  because there it sees gain from approximately the entire length of the BEC. At this point, atoms are transferred from $\psi_0$ to $\psi_2$ concurrently with the growth of $e_-$ and reduction in $e_+$. When the population in $\psi_0$ is sufficiently depleted at the input edge of the BEC, the process slows down, and the light field envelopes move towards the centre of the condensate, where $\abs{\psi_0}$ is still large, and the exchange between the two light fields continues; this is the time shown in Fig.~\ref{fig:3}. At this time, $\abs{\psi_0}$ grows again at the input end of the condensate, driven there by the destruction of $e_-$ photons generated further inside the sample, and $\psi_2$ atoms. In this way, the back-scattered photon flux out the input end of the condensate stops, and the first superradiant pulse has been emitted. The basis of `ringing' behaviour - the emission of subsequent SR pulses - is merely a repetition of the dynamics described above. Furthermore, a fascinating implication of the above dynamics, visible in Fig.~\ref{fig:3} (c) and (d), is that at times the light intensity within the BEC is higher than outside - the interaction leads to the formation of an optical resonator, where  partially reflecting mirrors are formed by the density modulation due to the interference of stationary and recoiling matter-waves. These Bragg gratings are centered where $\psi_0$ and $\psi_2$ cross.

The dynamics in the low detuning case is similar, but with two significant differences. Due to the increased light scattering cross-section at low detunings, the pump light is significantly depleted in its passage through the BEC, as is evident in Fig.~\ref{fig:3}(d). Accordingly, the build up of back-scattered light is even more localized at the input end of the BEC, and the scattering of $\psi_0$ to $\psi_2$ atoms is limited to a very narrow region of atoms. This narrowness is reflected at later times in the length of the `resonator', as is evident in Fig.~\ref{fig:3}(b). Thus, the spatial dependence of the pump depletion and hence the size of the scattering region  leads to the reduced amplitude of the back-scattered pulses in Fig.~\ref{fig:2}.
The second main difference is that the $\psi_4$ mode becomes  significantly populated at low detunings, as can be seen in Fig.~\ref{fig:3}(b). This means that the first SR pulse is not stopped by a lack of $\psi_0$ atoms, but rather can continue due to the scattering from $\psi_2$ to $\psi_4$ via the destruction of a pump photon. This can be seen in Fig.~\ref{fig:3}(d), where there is more light `leaking' from the resonator compared to Fig.~\ref{fig:3}(c); and in Fig.~\ref{fig:2}(c), where the low detuning pulses are broader, and show a secondary peak soon after the first. In general, the dynamics become complicated as the number of significantly atomic orders and hence number of timescales in the problem ($\propto 1/\omega_m$) increases.

In summary, we have studied superradiant scattering of light
by Bose condensed atomic samples in new parameter regimes
and find very good agreement between experimentally detected light
pulses and simulations. The simulations take into account the build-up of
longitudinal spatial structure inside the sample of both the light and matter waves.
The results help to identify a suitable parameter regime for the experimental study of correlations between light and atoms that arise from superradiance.
\begin{acknowledgments}
A. Hilliard would like to thank D. Witthaut for his help with the simulations.
This work is supported by the DGF center QUANTOP and
EU-projects EMALI and QAP.
\end{acknowledgments}
\vspace{-0.5cm}
%\bibliography{C:/home/Thesis/Biblio}

\end{document}